\documentclass[%
 aip,
 amsmath,amssymb,
 reprint,%
]{revtex4-1}
\usepackage{graphicx}
\usepackage{dcolumn}
\usepackage{amssymb,amsfonts,amsmath}
\usepackage{color}

\begin{document}
\title{Density scaling in the mechanics of a disordered mechanical meta-material} 
\author{Daniel Rayneau-Kirkhope}
\affiliation{Department of Applied Physics, Aalto University, P.O. Box 11100, FI-00076 Aalto, Finland}
\author{Silvia Bonfanti}
\affiliation{Center for Complexity and Biosystems, Department of Physics
University of Milano, via Celoria 16, 20133 Milano, Italy}
\author{Stefano Zapperi}
\affiliation{Center for Complexity and Biosystems, Department of Physics
University of Milano, via Celoria 16, 20133 Milano, Italy}
\affiliation{CNR - Consiglio Nazionale delle Ricerche, Istituto di Chimica della Materia Condensata e di Tecnologie per l'Energia, Via R. Cozzi 53, 20125 Milano, Italy}
\begin{abstract}
Nature provides examples of self-assemble lightweight disordered network structures 
with remarkable mechanical properties which are desirable for many applications purposes 
but challenging to reproduce artificially. 
Previous experimental and computational studies investigated the mechanical responses of
random network structures focusing on topological and geometrical aspects in terms of variable connectivity or 
probability to place beam elements. 
However for practical purposes an ambitious challenge is to design new materials with the possibility to 
taylor their mechanical features such as stiffness. 
Here, we design a two dimensional disordered mechanical meta-material exhibiting unconventional stiffness-density scaling in the regime where both bending and stretching are relevant for deformation. 
In this regime, the mechanical meta-material covers a wide interval of the Young modulus - density plane, simultaneously exhibiting high critical stress and critical strain.
Our results, supported by finite element simulations, provides the guiding principles to design on demand disordered metamaterials, bridging the gap between artificial and naturally occurring materials.
\end{abstract}
\maketitle
The mechanical properties of lattice materials of many types, such as foams~\cite{ashby2006properties}, 
cellular solids~\cite{gibson1999cellular,wadley2006multifunctional,ashby1983mechanical}, 
microlattices~\cite{fleck2010micro}, trusses~\cite{deshpande2001effective}, made of connected elements, have been intensively studied mainly due to their lightweight structures and remarkable mechanical properties~\cite{gibson2005biomechanics,phani2017dynamics}. 
The high strength to weight ratio of bone \cite{bone1, bone2} or balsa~\cite{zheng2014ultralight}, the elastic properties of spider silk \cite{silk1, silk2} and the fracture resistance of nacre \cite{nacre1, mayer2005rigid} are just few of the naturally occurring structures
that derive their mechanical properties from their underlying geometry.
These types of materials attract growing interest in many fields ranging from commercial products such as those related to food industry, 
to architectural applications such as energy absorption and management~\cite{ashby2006properties} and 
in modern technologies where their geometrical features are exploited to achieve a myriade of performances. 

In the recent years composite structures started to be rationally architected with the aim to achieve targeted properties~\cite{aux2}, and 
emerging significant breakthroughs~\cite{frenzel2017three} are favoured alongside with the advances in digital manufacturing technologies i.e. 3D printing and automated assembly.
Artificially designed materials are recently termed meta-materials~\cite{buckmann2012tailored,paulose2015selective,energy2,order1,coulais2017extension}, specifically, mechanical meta-materials indicate a class of structures whose mechanical properties are a consequence of their underlying geometry rather than their constituent material \cite{meta_mat1, meta_mat2, meta_mat}. Through the prudent choice of a meta-materials underlying architecture, it is possible to create geometries whose structural performance far exceeds that of the material from which it is made \cite{meta_mat2}. 
These structures can be designed to exhibit a wide range of beneficial properties, including high strength to weight ratio \cite{stiff-strong1, stiff-strong2}, auxetic behaviour \cite{aux1, aux2,aux3,aux4}, energy trapping \cite{energy1, energy2} and fracture resistance \cite{fracture1, fracture2}, among many others \cite{meta_mat1}. 

Typically artificil meta-materials are designed with a single motif repeated periodically throughout the material \cite{order1}, in contrast with the disordered structures encountered in self-assembled natural objects \cite{nature1, nature2}. 
Random elastic networks are widely studied~\cite{feng1984percolation,wilhelm2003elasticity,yan2017edge}, 
but there are still large gaps between theoretical investigations and practical purposes such as real solid materials. 
Mechanical meta-materials with increasing degree of disorder are just emerging~\cite{dis1,dis2,dis3,yang2018theoretical,reid2018auxetic}.
Whether disorder is merely the price to pay for self-assembly or whether it provides some advantage is still an open issue. %

The structural behaviour of lattice based meta-materials, made up of beams with non-hinged nodes, is generally divided into two regimes: stretching-dominated and bending-dominated \cite{Ashby2006}. Which of these two regimes a lattice falls into is dependent on its connectivity: When a lattice's connectivity is less that is required for rigidity in its rigid link freely hinged analogous system, the lattice will exhibit bending dominated behaviour, for lattices with connectivity above this threshold, stretching dominated behaviour will be observed \cite{Ashby2006}. The two classes of lattice exhibit dissimilar mechanical properties. 
For example, the relative stiffness of a two-dimensional lattice ($\tilde{Y}/Y$) is related to its relative density ($\tilde{\rho}/\rho$) via the expression
\begin{equation}
\frac{\tilde{Y}}{Y} \sim \left(\frac{\tilde{\rho}}{\rho}\right)^n\label{stiff}
\end{equation}
where, in two-dimensions, $n$ is one or three for stretching-dominated and bending dominated architectures respectively \cite{Stiff_aux} (in $d=3$, $n$ takes the values 1 or 2 for stretching and bending dominated architectures respectively \cite{Ashby2006}). Due to this scaling behavior, stretching-dominated lattices of low relative density are stiffer than their bending-dominated counterparts.  In the above expressions $Y$ and $\rho$ are the Young's modulus and density of the construction material respectively, while $\tilde{X}$ denotes the property $X$ of the meta-material. Density dependent scaling laws similar to the one reported in Eq. \ref{stiff} can be 
also be established for other mechanical properties, such as buckling or strength, or even electrical properties \cite{Ashby2006}.

Better understanding of stochastic lattice based structures has importance for a range of applications from performance of lightweight structures \cite{si,worsley,wang} to early diagnosis of diseases. 
For example, osteoporosis, a disease with major socio-economic consequence, is linked both with changes in bone mineral density (BMD) and also changes in connectivity of the trabecular bone latticework \cite{osteoporosis}. Improving the comprehension of the implications of density and connectivity in stochastic lattices is thus a key step in the early diagnosis of this disease where currently BMD is the principle diagnostic tool \cite{osteoporosis}. 

Inspired by naturally designed disordered structures, in this letter, we consider the linear response and elastic limit of a disordered lattice made up of beam elements. In contrast to periodic structures previously investigated, 
disordered lattices exhibit an intermediate regime characterized by an alternative relationship linking relative stiffness and relative density of the lattice. Furthermore, we establish the effect of diluting the lattice on the buckling resistance of the meta-material and show that in the intermediate regime, a novel scaling law linking critical strain (strain at buckling) and relative density exists. This mixed regime exhibits many advantageous properties. At low densities the lattice far exceeds the bending dominated lattice in terms of stiffness. At higher densities, the mixed regime simultaneously shows high buckling load and critical strain. 
With increasing manufacturing freedom to design structures on the micro and nano scale  \cite{nano1, nano2}, and the possibility of self assembly of complex architectures \cite{self1, self2}, the theoretical work presented here is of increasing technological relevance.

To investigate the density scaling of disordered meta-materials, we consider a triangular lattice whereby each linkage element (beam of length $L$ and thickness $t$) is placed with a given probability $p$. Equivalently, we can create the same structures by removing the beam with the probability $(1-p)$ from the perfect lattice. The designed geometries are subsequently 
deformed under a strain in the negative $y$ direction (in all cases, the imposed strain is less than the critical strain).  The thickness of the beams was varied in order to vary the relative density. 
Each beam element is made of a linear elastic material with Young's modulus $Y$ and Poisson's ratio $\nu$, we define the aspect ratio of the beam to be $a = L/t$. 
From the resultant architecture we take the largest connected component and investigate the  mechanical properties resultant meta-material. 
The system size is such that the entire lattice measures $N_x l$ horizontally, and $\sqrt{3} l N_y/2$ vertically.

All simulations presented here have been performed using COMSOL Multiphysics \cite{COMSOL} and COMSOL with MATLAB \cite{COMSOL} in $d=2$ with a plane strain approximation through use of the structural mechanics module.  All studies assume a linear elastic material with Young's modulus of 50MPa and Poisson's ratio of 0.3; the length of the beams used was 0.01m.
Mesh refinement studies were undertaken to ensure convergence of results. Results pertaining to stiffness have been obtained using Euler-Bernoulli beam elements using the in-built ``stationary studies'' study (a quasi-static solver). Here, a small displacement was applied to the upper boundary, while the lower boundary was fixed in place inducing a reduction in total height of the structure, the remaining boundaries were left free; the reaction force required to induce this displacement was then calculated and from this, an effective Young's modulus was calculated. For studies into buckling strength and critical strain the lower boundary was again fixed, while the upper boundary had an applied force or displacement imposed (for buckling stress and strain respectively), the remaining boundaries were left free. For these studies, the solid mechanics module used alongside the in-built linear buckling solvers (linear eigenvalue solver). For studies using the solid mechanics module, mesh density required were highly dependent on the aspect ratio of the beams, approximately 200-1500 elements were used per unit cell.   

\begin{figure}[th]
\begin{center}
\includegraphics[width=8.1cm]{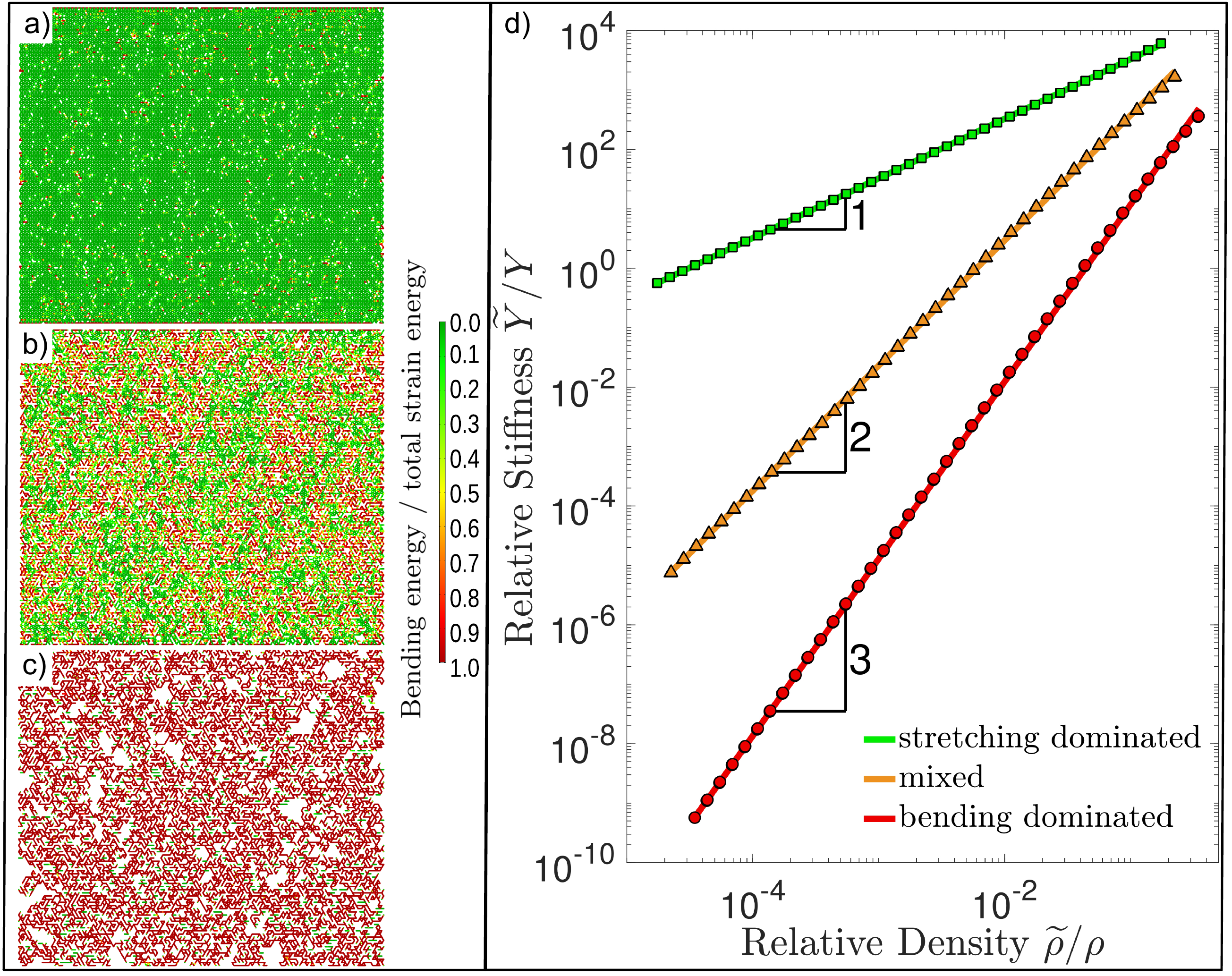}
\end{center}
\caption{Different scaling regimes of stiffness vs relative density. Examples of typical 128$\times$128 node networks with different bond placing probability a)~$p =0.95$, b)~$p = 0.642$ and c)~$p=0.4$ with a small strain imposed in the $y$ direction. The colors show the proportion of the local strain energy that is stored as bending energy with red indicating bending energy dominates, yellows showing mixed and green indicating stretching behavior. d)~Stretching ($n=1$), mixed ($n=2$) and bending ($n=3$) dominated lattices are characterised by power law behaviors, fitting of the data shown gives scaling of $n =$ 2.97 $\pm 0.02$, 2.1$\pm 0.05$ and 1.00$\pm 0.01$ for bending, mixed and stretching respectively.} 
\label{geometries}
\end{figure}

Figure \ref{geometries}a) shows an example of three geometries for $p=0.4,0.642,0.95$ deformed along negative $y$ direction. Beams are colored according to the proportion of bending energy with respect to the total \cite{Rayneau-Kirkhope2018}. 
Increasing $p$, the designed random meta-materials cover different ranges from bending to stretching behavior. 
Figure~\ref{geometries}b shows that for a dilute triangular lattice, close to the transition from 
the stretching dominated to the bending dominated regime ($p = p_{\mbox{\tiny T}}$, in this work $p_{\mbox{\tiny{T}}}$ is taken to be 0.642 inline with previous estimates \cite{Roux, Hansen, Arbabi}) an intermediate scaling regime is observed where contributions to stiffness from both stretching and bending cannot be ignored. This results in a new exponent $n \simeq 2$ that lies between the  value $n=1$ expected for stretching dominated elasticity and $n=3$ for bending dominated
elasticity. 

\begin{figure}[h]
\begin{center}
\includegraphics[width=8.1cm]{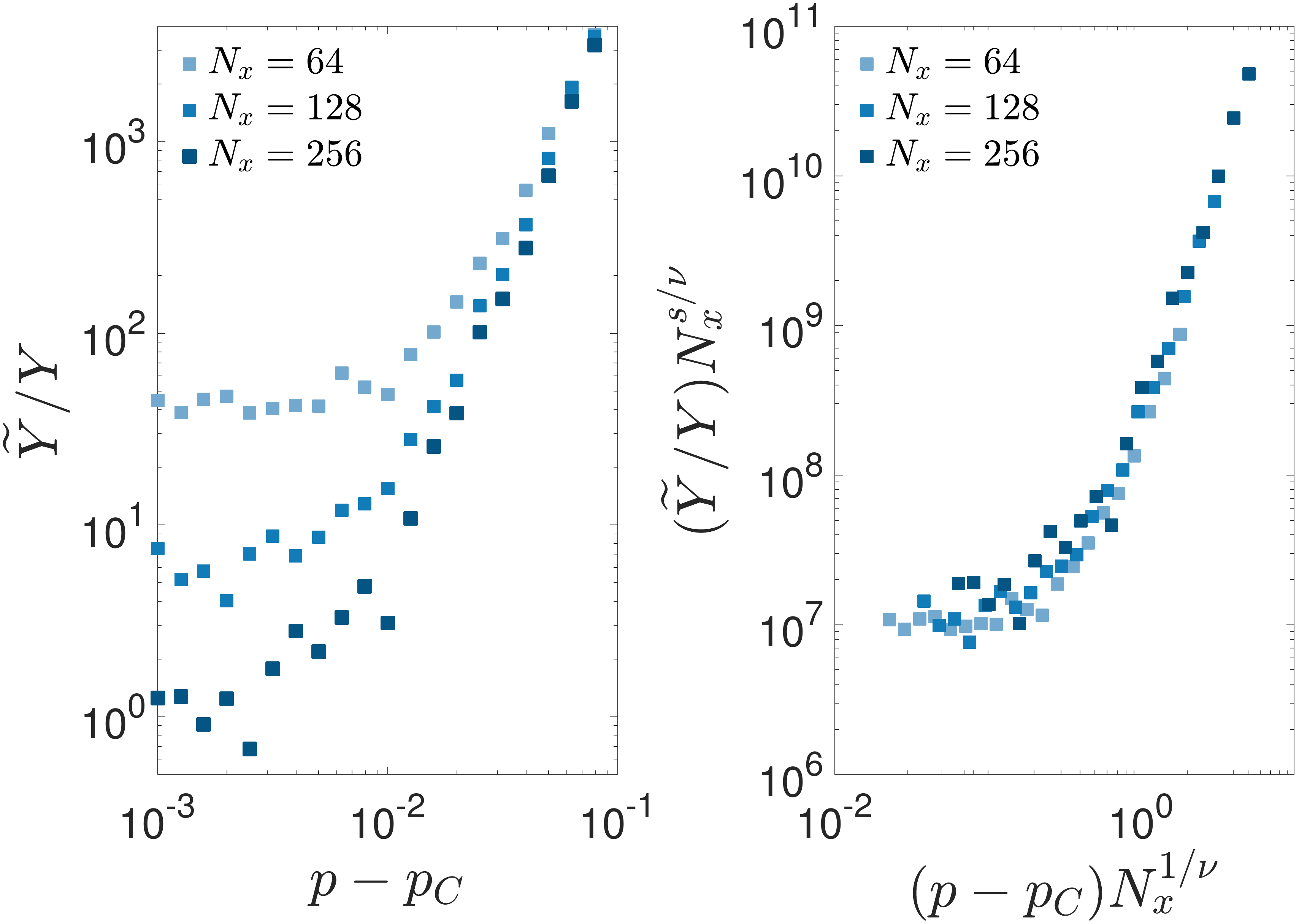}
\end{center}
\caption{a)~The Young modulus $\tilde{Y}$ as a function of $p-p_c$ for different values of 
$N_x$. The data shown is for beams with an aspect ratio of 16. Each data point shown is an average over a number of simulations $N_s$ such that the product of $N_s$ and the number of elements in the perfect system is greater than $10^6$.  \label{fig:pc_scaling} 
b)~Rescaling of the data in a) according to Eq. \ref{eq:FSS}.}
\end{figure}

It is noted that according to effective-medium theory (EMT), in this regime the contribution to stiffness will follow \cite{EMT}:
\begin{equation}
\tilde{Y} \sim \mu^{(1-x)}\kappa^x,
\end{equation}
where $\mu$ and $\kappa$ are the stretching and bending stiffness of the lattice elements, respectively. EMT predicts that $x=0.5$ in $d=2$ and $x=0.4$ in $d=3$ \cite{EMT}. 
For a lattice made up of beam elements of thickness $t$ and length $L$, we note that 
$\mu \sim (t/L)^q$ and $\kappa \sim (t/L)^r$,
where in $d = 2, q =1$ and $r = 3$, while for $d=3, q = 2$ and $r=4$.
Since $\tilde{\rho}/\rho \propto (t/L)$ in $d=2$ and $\tilde{\rho}/\rho \propto (t/L)^2$ in 
$d=3$
within the intermediate regime, we expect to observe the scaling,
$\tilde{Y}/{Y} \sim (\tilde{\rho}\rho)^n,$ with $n=2$ in $d=2$ and $n=1.4$ in $d=3$. 
This scaling is plotted against the results of simulation for probabilities close to the critical probability $p_{\mbox{\tiny{T}}}$ in \ref{geometries}b. 
This intermediate scaling close to the isostatic point is reminiscent observed in stochastic fibre networks where stretching and bending energies can be varied arbitrarily. 
In contrast, in this work the geometric parameters of the beam elements set the ratio of bending to stretching energies in the beam elements \cite{Timo}, thus here we observe a dependence of the stiffness of the meta-material on its relative density.

In contrast to the behaviour of fibre networks which exhibit a ``bending rigidity percolation'' probability which can be calculated through Maxwell counting and mean field arguments \cite{Maxwell_argument}, here we observe that connectivity of the upper and lower boundaries is sufficient to yield a non-zero stiffness in a stochastic meta-material constructed from beam elements. Thus we see non-zero effective Young's modulus for probabilities greater than the 
geometric percolation probability $p_c$ that for the triangular lattice
is given by \cite{Sykes} $p_c = 2\sin(\pi/18)$.

For $p-p_c \ll 1$ we find that the Young's modulus obeys standard
finite size scaling expected for percolation 
\begin{equation}
\tilde{Y} \sim N_x^{-s/\nu} f((p - p_c) N_x^{1/\nu})
\label{eq:FSS}
\end{equation}
where $\nu=4/3$ is the correlation length exponent, while $s$ is the elasticity exponent
that can be expressed in terms of the conductivity exponent $t\simeq 0.98\nu $
\cite{grassberger} as $s=t+2\nu \simeq 4$ \cite{Sahimi}. Eq. \ref{eq:FSS} is verified by the data collapse reported in Fig. \ref{fig:pc_scaling} for three different lattice size with $N_x$=64, 128, 256.

\begin{figure}[th]
\begin{center}
\includegraphics[width=7.9cm]{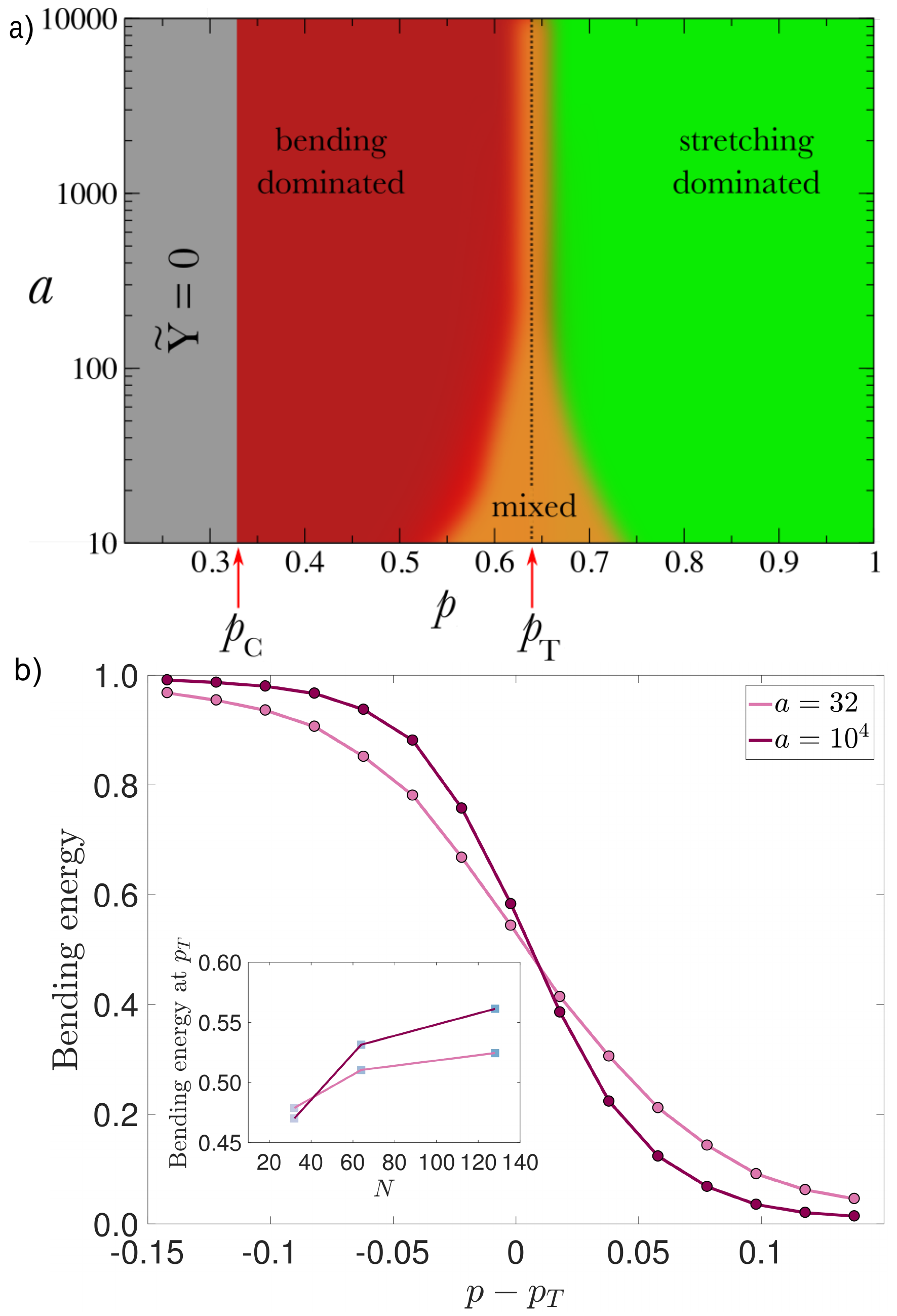}
\end{center}
\caption{a) The parameter space $(p,a)$ and the observed regime of stretching and bending dominated alongside a mixed regime. b) The transition from bending (low $p$) to stretching (high $p$) dominated regimes for system size $N_x = 128$ for systems made up of beams with aspect ratio $32$ and $10^4$, a sharper transition is seen for higher aspect ratio of beams. The inset reports the variation of the bending energy as a function of three system sizes, 
$N_x = 32,64,128$.
Figure shows averages over a number of simulations $N_s$ such that to product of the total number of elements in the perfect system ($p=1$) and $N_s$ is greater than $2\times 10^5$.\label{fig:phases}}
\end{figure}

\begin{figure}[h]
\begin{center}
\includegraphics[width=8.1cm]{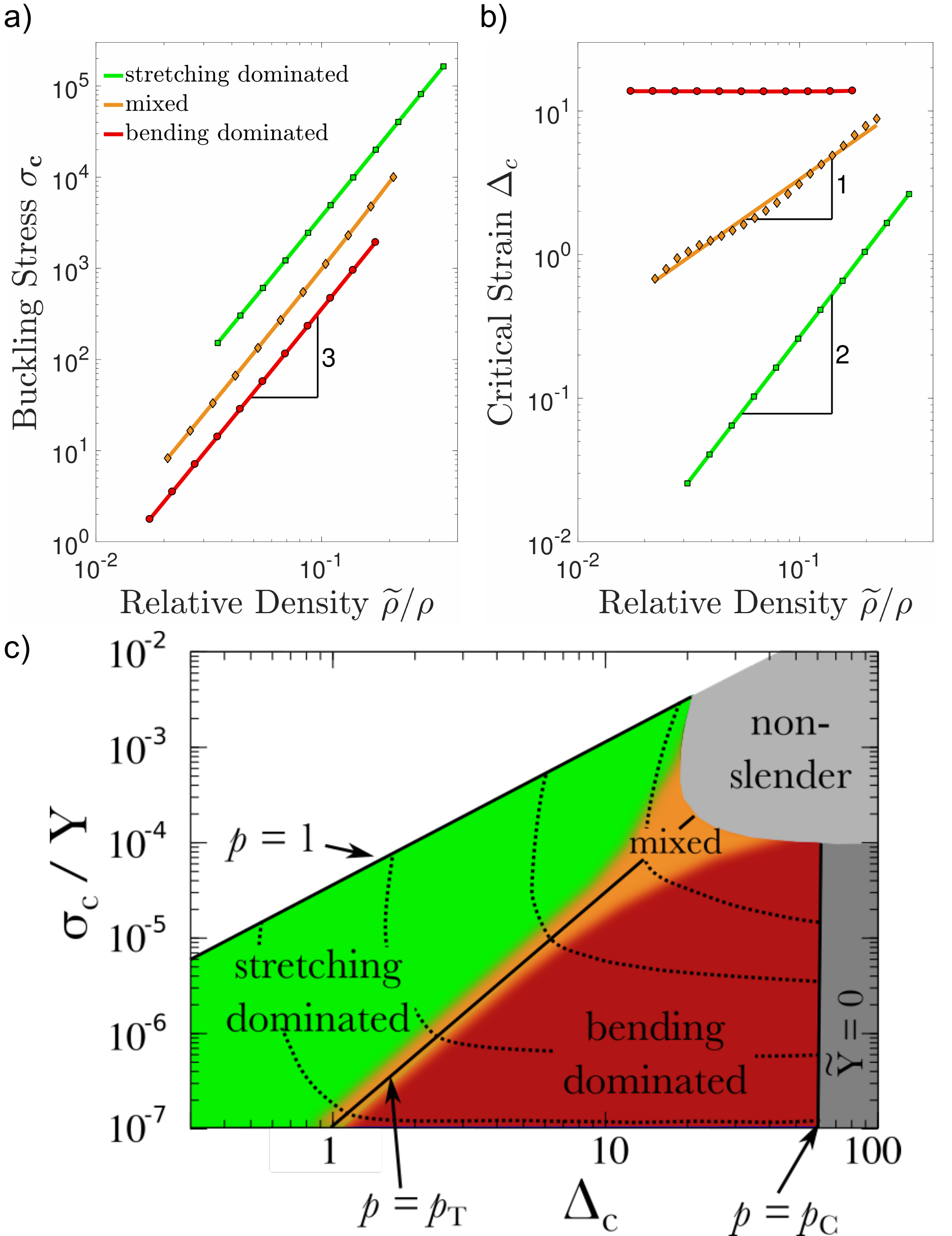}
\end{center}
\caption{a) The critical stress for lattices in the bending, mixed and stretching regimes. All regimes show the same scaling of $\sigma_c$ vs relative density as predicted by Eq.~(\ref{stress}). b)~The critical strain for the stochastic metamaterial for the different regimes, here the scaling is dependent on the regime, and is predicted by Eq.~(\ref{strain}). c)~Stretching dominated regimes leads to high critical stress with low critical strain, bending dominated lattices exhibit high critical strain but low critical stress; Lattices with significant energy contributions from both bending and stretching simultaneously exhibit both high stress and strain at failure. The data shown is for a system size of $N_x = N_y = 20$.
\label{strain_vs_density}}
\end{figure}

We summarize these finding in figure \ref{fig:phases} where we show the onset of non-zero stiffness at $p_c$, and the region of $(p,a)$-space for which the bending, stretching and mixed regimes are observed. With decreasing slenderness of beams (high relative densities), a broader range of $p$ leads to lattices where both stretching and bending energies must be considered.

In the stochastic metamaterial made up of slender beam elements, elastic instability will occur when the loading reaches a critical value. It is expected that in $d = 2$ or $d=3$, the critical value will scale as 
\begin{equation}
\sigma_c \sim F_c/L^{d-1} \sim YI/L^{d+1}\label{buck_scaling}
\end{equation}
 where $\sigma_c$ is the critical stress, $F_c$ is the Euler buckling force of a slender beam, $I$ and $L$ are the second moment of area and length of the component beams respectively, and $Y$ is the Young's modulus of the construction material. 
Thus, through variation of the aspect ratio of the component beams with a fixed $p$, we expect a scaling relationship between critical strain and density given by, 
\begin{equation}
\sigma_c \sim \left(\frac{\tilde{\rho}}{\rho}\right)^{m},\label{stress}
\end{equation}
where, in $m=3$ in $d=2$ and $m =2$ in $d=3$ \cite{Ashby2006}.  This is confirmed in figure \ref{strain_vs_density} where the scaling of $m=3$ is found irrespective of the regime (stretching, bending or mixed) to which the lattice belongs. 
As a result of this scaling, we calculate the dependence of critical strain on the relative density of the lattice. 
It is predicted here that $\Delta_c$ will follow the scaling, 
\begin{equation}
\Delta_c \sim \left(\frac{\tilde{\rho}}{\rho}\right)^{(m-n)}\label{strain}
\end{equation}
where in $d=2$, $m = 3$ irrespective of the regime, and $n$, as in Eq.~(\ref{stiff}), takes the values 1, 2 or 3 when the lattice is stretching dominated, mixed, or bending dominated respectively (in $d=3$, $m=2$ irrespective of the regime, and $n$ is expected to take the value 1, 1.4 or 2 for stretching dominated, mixed or bending dominated respectively); 
this prediction for $d=2$ is confirmed in figure (\ref{strain_vs_density}). 
We thus find that coupling high relative density and high relative strength is achieved through the use of the mixed deformation mode, that is $p \approx p_{\mbox{\tiny{T}}}$, this is shown in figure \ref{strain_vs_density}.

Lattice based meta-materials, made up of beam elements, have been traditionally classed as either bending or stretching dominated depending on their connectivity \cite{Ashby2006}. 
It is notable that recent experimental work has estabilished that some 3 dimensional lightweight nanolattices exhibit density scaling that is inconsistent with these classifications \cite{Meza1, Meza2, Zhang}. 
Here we have established, through theory and simulation, the existence of a third class of lattice architecture where both bending and stretching energies must be considered. 
We have established that such architectures lead to meta-materials with high critical stress and critical strain. 
We plan to verify our predictions for the relevant scaling relationships for three dimensional geometries, which would be computationally more demanding. 

It is notable that in this stochastic methodology generates structures that are closer in design to their naturally occurring counterparts. 
These naturally occurring structures exhibit high flaw tolerance and insensitivity to perturbations \cite{nature2}. 
Though this flaw tolerance is linked with optimisation of the materials geometry and hierarchy \cite{geo, nature2}, 
it remains a plausible possibility that flaw tolerance is aided by disorder. 
Such a link between randomness and robustness has been previously been found in machine resilience \cite{MR}, algorithm design \cite{AD} and interdependent lattice networks \cite{ILN}. 
Extending this link to structural mechanics has huge potential for the design of functional robust structures.

\vspace{5mm}

D.~R.-K. would like to acknowledge funding support from the Academy of Finland postdoctoral grant program.  S.Z acknowledges support from the Academy of Finland FiDiPro progam, project 13282993. S. Z. and S. B. are supported by the project DISORDER 
funded by the Ministero degli Affari Esteri e della Cooperazione Internazionale.


\end{document}